\documentstyle[12pt]{article}

\begin{document}

\title{On the application of one M.G.Krein's result to the spectral analysis of
Sturm-Liouville operators. }
\date{April, 2000 }
\author{S.A.Denisov \thanks{
Moscow State University. Mathematical Subject Classification 34B24, 34L20.
The work was partially supported by RFBR Grant N00-15-96104.}}
\maketitle

\begin{abstract}
Discovered by M.G.Krein analogy between polynomials orthogonal on the unit
circle and generalized eigenfunctions of certain differential systems is
used to obtain some new results in spectral analysis of Sturm-Liouville
operators.
\end{abstract}

{\bf Section A.}

In this section we remind some results obtained by M.G.Krein in his famous
article \cite{Kr}. In this paper author develops the ''theory of
polinomials, orthogonal on the positive half-line''. And this polinoms are
constructed from exponents rather than from the powers of independent
variable. It's well known that there are many ways to construct the system
of orthogonal polinomials on the unit circle. One of them is to start from
the moments matrix. This way was chosen by M.G.Krein to obtain his results
for positive half-line.

Let's assume that $H(t)=\overline{H(-t)}$ - function summable on each
segment $(-r,r)$.

{\bf Proposition.} {\it If for any continuous }$\varphi (t)${\it \ the
following inequaliy holds }

\begin{equation}
\int\limits_0^r\left| \varphi (s)\right|
^2ds+\int\limits_0^r\int\limits_0^rH(t-s)\varphi (t)\overline{\varphi (s)}%
dtds\geq 0  \label{asr1}
\end{equation}

{\it for each }$r>0${\it , so, and in this case only, there exists the
non-decreasing function }$\sigma (\lambda )${\it \ }$(\lambda \in R,\sigma
(0)=0,\sigma (\lambda -0)=\sigma (\lambda )${\it \ }$)${\it , such that } 
\begin{eqnarray*}
\int\limits_{-\infty }^\infty \frac{d\sigma (\lambda )}{1+\lambda ^2}
&<&\infty \\
\int\limits_0^t(t-s)H(s)ds &=&\int\limits_{-\infty }^\infty (1+\frac{%
i\lambda t}{1+\lambda ^2}-e^{i\lambda t})\frac{d\sigma (\lambda )}{\lambda ^2%
}+(i\gamma -\frac{sign(t)}2)t \\
&&
\end{eqnarray*}
{\it \ }

{\it where }$\gamma ${\it \ is real constant.}

If in addition we presume that the equality in (\ref{asr1}) is possible for $%
\varphi =0$ only then the Hermit kernel $H(t-s),\ (0\leq t,s\leq r)$ has
Hermit resolvent $\Gamma _r(s,t)=\overline{\Gamma _r(s,t)}$ that satisfies
the relation

\[
\Gamma _r(t,s)+\int\limits_0^rH(t-u)\Gamma _r(u,s)du=H(t-s)\ (0\leq s,t\leq
r) 
\]

The continuous analogues of polinomials orthogonal on the unit circle are
defined by the formula

\begin{eqnarray*}
P(r,\lambda ) &=&e^{i\lambda r}(1-\int\limits_0^r\Gamma _r(s,0)e^{-i\lambda
s}ds), \\
P_{*}(r,\lambda ) &=&1-\int\limits_0^r\Gamma _r(0,s)e^{i\lambda s}ds,\ r\geq
0.
\end{eqnarray*}

Using the well known properties of resolvents we obtain the following system

\begin{equation}
\begin{array}{ccc}
\frac{dP(r,\lambda )}{dr} & = & i\lambda P(r,\lambda )-\overline{A(r)}%
P_{*}(r,\lambda ), \\ 
\frac{dP_{*}(r,\lambda )}{dr} & = & -A(r)P(r,\lambda ),
\end{array}
\label{sys3}
\end{equation}

where $A(r)=\Gamma _r(0,r).$

{\bf Proposition.} {\it For each finite }$f(x)\in L^2(R^{+})${\it \ we have
the following equality}

\[
\left\| f\right\| _2^2=\int\limits_{-\infty }^\infty \left| F_P(\lambda
)\right| ^2d\sigma (\lambda ),{\rm where\, } F_P(\lambda
)=\int\limits_0^\infty f(r)P(r,\lambda )dr. 
\]

Consequently we have the isometric mapping $U_P$ from $L^2(R^{+})$ into $%
L^2(\sigma ,R).$

{\bf Theorem.} {\it The mapping }$U_P${\it \ is unitary if and only if the
following integral diverges \\(equals to }$-\infty ${\it \ )}

\begin{equation}
\int\limits_{-\infty }^\infty \frac{\ln \sigma ^{^{\prime }}(\lambda )}{%
1+\lambda ^2}d\lambda .  \label{int2}
\end{equation}

{\bf Theorem.} {\it \ The following statements are equivalent}

{\it (1) The integral {\rm (\ref{int2})} is finite.}

{\it (2) At least for some }$\lambda $, $\Im \lambda >0$ {\it the integral } 
\begin{equation}
\int\limits_0^\infty \left| P(r,\lambda )\right| ^2dr  \label{int3}
\end{equation}

{\it converges.}

{\it (3) At least for some }$\lambda ${\it \ (}$\Im \lambda >0${\it \ ) the
function }$P_{*}(r,\lambda )${\it \ is bounded.}

{\it (4) On any compact set in the open upper half-plane integral {\rm (\ref
{int3})} converges uniformly. That is equivalent to the existence of uniform
limit }$\Pi (\lambda )=\lim_{r\rightarrow \infty }P_{*}(r,\lambda ).${\it \ }

It's easy to verify that in cases $A(r)\in L^1(R^{+}),\ A(r)\in L^2(R^{+})$
the conditions (1)-(4) are satisfied. What is more, in the first case
measure $\sigma $ is continuously differentiable with certain estimates for
its derivative.

Consider $E(r,\lambda )=e^{-i\lambda r}P(2r,\lambda )=\Phi (r,\lambda
)+i\Psi (r,\lambda ).$ Let $E(-r,\lambda )=\overline{E(r,\lambda )}=\Phi
(r,\lambda )-i\Psi (r,\lambda ).$ From (\ref{sys3}) we infer that 
\begin{eqnarray*}
\frac{d\Phi }{dr} &=&-\lambda \Psi -a(r)\Phi +b(r)\Psi ,\ \Phi (0,\lambda
)=1; \\
\frac{d\Psi }{dr} &=&\lambda \Phi +b(r)\Phi +a(r)\Psi ,\ \Psi (0,\lambda )=0.
\end{eqnarray*}

where $a(r)=2\Re A(2r),\ b(r)=2\Im A(2r).$

{\bf Proposition.} {\it The mapping }$U_E:f(r)\rightarrow F_E(\lambda
)=\int\limits_{-\infty }^\infty f(r)E(r,\lambda )dr,${\it \ defined on the
finite functions }$f(r)\in L^2(R),${\it \ generates the unitary operator
from }$L^2(R)${\it \ onto }$L^2(\sigma ,R)$. 

The trivial case $a=b=0$ yields $E(r,\lambda )=e^{i\lambda r},\ \Psi
(r,\lambda )=\sin (r\lambda ),\ \Phi (r,\lambda )=\cos (r\lambda ).$

In case when $H(t)$ is real, the function $\sigma (\lambda )$ is odd.
Consequently $b(r)=0.$ Assuming that $H(t)$ is absolutely continuous, we
have that $\Phi $ and $\Psi $ are solutions of the equations

\begin{equation}
\begin{array}{ccc}
\Psi ^{^{\prime \prime }}-q\Psi +\lambda ^2\Psi =0, & \Psi (0)=0, & \Psi
^{^{\prime }}(0)=\lambda ; \\ 
\Phi ^{^{\prime \prime }}-q_1\Phi +\lambda ^2\Phi =0, & \Phi (0)=1, & \Phi
^{^{\prime }}(0)+a(0)\Phi (0)=0,
\end{array}
\label{bis}
\end{equation}

where $q_1(x)=a^2(x)-a^{^{\prime }}(x)$ and $q(x)=a^2(x)+a^{^{\prime }}(x).$%
\vspace{1cm}

{\bf Section B.}

Consider the Sturm-Liouville operator on the half-line with Dirichlet
boundary condition at zero 
\begin{equation}
l(u)=-u^{^{\prime \prime }}+qu,\ u(0)=0.  \label{e1}
\end{equation}

Let's assume that real-valued $q(x)$ admits the following representation $%
q(x)=a^2(x)+a^{^{\prime }}(x)$ where $a(x)$ is absolutely continuous
function on the half-line. That means that $a(x)$ is the solution of the
Ricatti equation.

Consider also the corresponding differential Dirac-type system (see \cite{LS}
or \cite{Atk})

\begin{equation}
\left\{ 
\begin{array}{cc}
\Phi ^{^{\prime }}(x,\lambda )= & -\lambda \Psi (x,\lambda )-a(x)\Phi
(x,\lambda ) \\ 
\Psi ^{^{\prime }}(x,\lambda )= & \lambda \Phi (x,\lambda )+a(x)\Psi
(x,\lambda )
\end{array}
\right. ,  \label{s1}
\end{equation}

where $\Phi (0,\lambda )=1,\Psi (0,\lambda )=0.$

From the result stated in Section A it follows that the spectral measure $%
\rho (\lambda )$ of problem (\ref{e1}) is connected with the spectral
measure $\hat{\sigma} (\lambda )$ of system (\ref{s1}) \footnote{%
See the definition of spectral measure for the differential system in \cite
{Atk}. Here the function $\hat{\sigma} (\lambda )$ is connected with $\sigma
(\lambda)$ from the section A by the relation: $\hat\sigma(\lambda
)=2\sigma(\lambda ).$} by the following relation

\begin{equation}
\rho (t)=2\int\limits_0^{\sqrt{t}}\alpha ^2d\hat{\sigma} (\alpha ).
\label{m1}
\end{equation}

From this and results stated in Section A we can infer one very simple but
significant corollary

{\bf Corollary. } {\it If }$q(x)${\it \ is real-valued function such that }

{\it 
\begin{equation}
\sup\limits_{x\in R}\int\limits_x^{x+1}\left| q(s)\right| ^2ds<\infty ,
\end{equation}
}

{\it the improper integral }$W(x)=\int\limits_x^\infty q(s)ds${\it \ exists
and satisfies the condition $W(x)\in L^2(R^{+}),$ then the absolutely
continuous part of spectrum of operator $H_h$, generated by differential
expression $l(u)=-u^{^{\prime \prime }}+qu$ and boundary condition$\
u(0)=hu^{^{\prime }}(0),\ (h\in R\cup \infty \ )$ fills the whole positive
half-line.}

Proof. Let's consider (\ref{s1}) with $a(x)=-W(x).$ The spectral measure of
system (\ref{s1}) with chosen $a(x)$ has the needed property. Consequently
the Sturm-Liouville operator with potential $q^{*}(x)=a^{^{\prime }}+a^2$
and Dirichlet boundary condition also has the a.c. component which fills the
whole positive half-line. But initial potential $q(x)$ differs from $q^*(x)$
by $L^1(R^{+})$ term only. Consequently, Kuroda's theorem \cite{Kur}
guarantees that operator $H_h$ has the needed property for $h=0$. But it means
that this statement is true for any $h$ since the essential support of a.c.
component doesn't depend on $h.$ That follows, for example, from the
subordinate solutions theory.$\Box$

In the next theorems of this section we will show how Krein's results will
help to establish asymptotics for generalized eigenfunctions and analyze the
spectrum of some Sturm-Liouville operators.

{\bf Theorem 1. } {\it If }$q(x)${\it \ is real-valued function such that }

{\it 
\begin{equation}
\sup\limits_{x\in R}\int\limits_x^{x+1} \min\{0,q(s)\}ds>-\infty,
\label{aus}
\end{equation}
}

{\it the improper integral }$W(x)=\int\limits_x^\infty q(s)ds${\it \ exists
and satisfies the condition} $\left| W(x)\right| \leq \frac \gamma {x+1},\
(0<x,0<\gamma <1/4),${\it \ then operator }$H${\it \ generated by 
{\rm (\ref{e1})}
is non-negative and }

\begin{equation}
\left| \int\limits_0^\infty \frac{\ln \rho ^{^{\prime }}(\lambda )}{\sqrt{%
\lambda }(1+\lambda )}d\lambda \right| <\infty.  \label{t1}
\end{equation}

{\it \ What is more for a.e. positive spectral parameter the generalized
eigenfunctions of differential expression }$l(u)=-u^{^{\prime \prime }}+qu$%
{\it \ has the following asymptotic }$u(x,\lambda ,\alpha )=C(\lambda
,\alpha )\sin (x\lambda +\varphi (\lambda ,\alpha ))+\overline{o}(1),\
u(0,\lambda ,\alpha )=\cos (\alpha ),\ u^{^{\prime }}(0,\lambda ,\alpha
)=\sin (\alpha ).${\it \ The spectrum of operator $H$, generated by (\ref{e1}%
), is purely absolutely continuous on the positive half-line}.

Proof.

We will separate the proof on two parts.\vspace{0.5cm}

{\ 1. Reducing to the system.}

Let's consider Ricatti equation $q(x)=a^2(x)+a^{^{\prime }}(x)$ . We will
find solution which is absolutely continuous, tends to zero at the infinity
and belongs to the class $L^2(R^{+})$. Integrating we will get the following
nonlinear integral equation $\int\limits_x^\infty a^2(s)ds-a(x)=W(x).$ Our
goal is to study operator $B$: $Bf(x)=\int\limits_x^\infty f^2(s)ds-W(x)$
that acts in complete metric space $\Omega $ of measurable functions which
admit the estimate $\left| g(x)\right| \leq \frac \ae {x+1},\ (\ae =\frac{1-%
\sqrt{1-4\gamma }}2\ ).$

The metric is introduced by the formula $\rho (g_1,g_2)=ess\sup_{x\geq
0}\left\{ (x+1)|g_1(x)-g_2(x)\right |\} .$ To use the contraction operators
principle one should verify that the following conditions hold

1. $B$ is acting from $\Omega $ to $\Omega $,

2. $B$ is contraction operator.

It is not difficult to show that the both conditions are satisfied. Really

\[
\left| Bf\right| \leq \frac \gamma {x+1}+\int\limits_x^\infty \frac{\ae ^2}{%
(s+1)^2}ds=\frac{\ae ^2+\gamma }{x+1}=\frac \ae {x+1}, 
\]
\[
\left| Bg_1-Bg_2\right| \leq \int\limits_x^\infty \left| g_1-g_2\right|
\left| g_1+g_2\right| ds\leq \rho (g_1,g_2)\int\limits_x^\infty \frac{2\ae }{%
(s+1)^2}ds=\frac{2\ae }{x+1}\rho (g_1,g_2), 
\]

that means $\rho (Bg_1,Bg_2)\leq 2\ae \rho (g_1,g_2)$ which implies the
contraction property since $2\ae<1$. Thus we have the single fixed point $%
a(x)\in \Omega $, so that $Ba=a.$ Certainly this function satisfies the
Ricatti equation as well. So it suffices to use Proposition and (\ref{m1})
to obtain (\ref{t1}).\vspace{0.5cm}

{\ 2. Absence of singular component.}

Consider the system (\ref{s1}). From Section A we know that function

\begin{equation}
P(x,\lambda )=\exp (i\lambda \frac x2)\left( \Phi (x/2,\lambda )+i\Psi
(x/2,\lambda )\right)  \label{red}
\end{equation}

satisfies the following system

\[
\left\{ 
\begin{array}{ccc}
\frac{dP}{dx} & = & i\lambda P-AP_{*}, \\ 
&  &  \\ 
\frac{dP_{*}}{dx} & = & -AP.
\end{array}
\right. 
\]

Where $P(0,\lambda )=P_{*}(0,\lambda )=1,$ and $A(x)=\frac 12a(\frac x2).$
Let's introduce the following function $Q(x)=e^{-i\lambda x}P(x).$ So we
will have 
\[
\left\{ 
\begin{array}{ccc}
\frac{dQ}{dx}= & -Ae^{-i\lambda x}P_{*}, &  \\ 
&  &  \\ 
\frac{dP_{*}}{dx}= & -Ae^{i\lambda x}Q. & 
\end{array}
\right. 
\]

$P_{*}(0,\lambda )=Q(0,\lambda )=1.$ It's easy to see that $Q=\overline{P_{*}%
}$. Consequently $Q(x)=1-\int\limits_0^xA(s)e^{-i\lambda s}\overline{Q(s)}%
ds; $ $\left| Q(x)\right| \leq 1+\int\limits_0^x\left| A(s)\right| \left|
Q(s)\right| ds.$ Gronuol Lemma yields the estimate $\left| Q(x)\right| \leq
\exp \left( \int\limits_0^x\left| A(s)\right| ds\right) \leq \left( \frac{x+2%
}2\right) ^\ae .$ From (\ref{bis}) it follows that $u(x,\lambda )=\frac{\Psi
(x,\lambda )}\lambda $ $(\lambda \neq 0)$ satisfies the conditions $%
-u^{^{\prime \prime }}+qu=\lambda ^2u,$ $u(0,\lambda )=0,$ $u^{^{\prime
}}(0,\lambda )=1$. From (\ref{red}) we have $\left| u(x,\lambda )\right|
\leq \frac{(x+1)^\ae}{2^{\ae} \lambda} ,$ $(\ae <1/2$ $).$ The similar
estimate can be proved for linear independent solution $v(x,\lambda )$ such
that $v(0,\lambda )=1,\ v^{^{\prime }}(0,\lambda )=0.$ Indeed it suffices to
consider second equation of (\ref{bis}) letting $q_1=q$, solve equation $%
q=b^2-b^{^{\prime }}$ and repeat the same arguments to obtain the desired
inequality for linear independent solution $w(x,\lambda)$ which satisfies
the condition $w^{^{\prime }}(0)+b(0)w(0)=0$. Since $v(x,\lambda)$ is linear
combination of $u(x,\lambda)$ and $w(x,\lambda)$ we have the needed
estimate. In the same way derivatives $u^{^{\prime }}, v^{^{\prime}}$ can be
estimated. It follows from the inequality $|Q^{^{\prime}}|=|A||Q|\leq \frac{%
\ae }{2^{\ae}(2+x)^{1-\ae}}$ and (\ref{red}). Consequently from the
constancy of Wronskian $W(u,v)$ and equivalence of $\int\limits_x^{x+1}
u^2(s,\lambda) ds$ and $\int\limits_x^{x+1} {u^{^{\prime}}}^2(s,\lambda) ds$ 
\footnote{%
It's the only place in the whole proof where we use the condition (\ref{aus}%
).} we get the inequalities

\begin{eqnarray*}
C_2(\lambda )x^{1-2\ae } &\leq &\int\limits_0^xu^2(s,\lambda )ds\leq
C_1(\lambda )x^{2\ae +1}, \\
C_2(\lambda )x^{1-2\ae } &\leq &\int\limits_0^xv^2(s,\lambda )ds\leq
C_1(\lambda )x^{2\ae +1},
\end{eqnarray*}
where the constants $C_1,\ C_2$ are positive. So we can find $\zeta >0$ so
that $\left( \int\limits_0^xu_1^2(s,\lambda )ds\right) \left(
\int\limits_0^xu_2^2(s,\lambda )ds\right) ^{-\zeta }\rightarrow \infty $ for
any two linearly independent solutions $u_1,u_2$ of equation from (\ref{e1}).

The refined subordinacy theory \cite{Rem1}, \cite{JL} yields that there is $%
\eta (\lambda )>0$ so that $\left( D_\eta \rho \right) (\lambda ^2)=%
\overline{\lim }_{\varepsilon \rightarrow 0}$ $\frac{\rho (\lambda
^2-\varepsilon ,\lambda ^2+\varepsilon )} {\left( 2\varepsilon \right) ^\eta
}=0.$ Consequently $\rho $ gives zero weight to every $\Omega \subset R^{+}$
with $\dim \Omega =0.$ On the other hand we will prove that $u_1,u_2$ might
be unbounded at the infinity only on the $\lambda $ set with the zero
Hausdorff dimension. Consider $Q(x,\lambda ):\ Q^{^{\prime }}=-Ae^{-i\lambda
x}\overline{Q(x)},\ Q(0)=1.$

So

\begin{eqnarray*}
Q(x) &=&1-\int\limits_0^xA(s)e^{-i\lambda s}\overline{Q(s)}%
ds=1+\int\limits_0^x\left( \int\limits_s^\infty A(\tau )e^{-i\lambda \tau
}d\tau \right) ^{^{\prime }}\overline{Q(s)}ds= \\
&&
\end{eqnarray*}

\begin{equation}
\left. 1+\left( \int\limits_s^\infty A(\tau )e^{-i\lambda \tau }d\tau
\right) \overline{Q(s)}\right|_{s=0}^{s=x}-\int\limits_0^x\left(
\int\limits_s^\infty A(\tau )e^{-i\lambda \tau }d\tau \right) \overline{%
Q^{^{\prime }}(s)}ds,  \label{der}
\end{equation}

where $\lambda $ is such that integral $\int\limits_0^\infty A(\tau
)e^{-i\lambda \tau }d\tau $ converges. The last term in (\ref{der}) can be
rewritten as $\int\limits_0^x\left( \int\limits_s^\infty A(\tau
)e^{-i\lambda \tau }d\tau \right) A(s)e^{i\lambda s}Q^{^{}}(s)ds.$ Finally
we get

\[
Q(x)=J(x)-\int\limits_0^x\left( \int\limits_s^\infty A(\tau )e^{-i\lambda
\tau }d\tau \right) A(s)e^{i\lambda s}{Q^{^{}}(s)}ds, 
\]

where $J(x)=C(\lambda)+o(1)\overline{Q(x)}$. From argument used in \cite
{Rem2} (Theorem 1.3, 1.4 ) it follows that the set $\Xi $ of $\lambda $ for
which $\left( \int\limits_s^\infty A(\tau )e^{i\lambda \tau }d\tau \right)
A(s)e^{-i\lambda x}\notin L^1(R^{+})$ has zero Hausdorff dimension. But one
can easily verify that this fact leads to the boundedness of $Q(x)$ at the
infinity for $\lambda \notin \Xi .$ From formula (\ref{red}) it follows that
generalized eigenfunctions $u(x,\lambda )$ which correspond to the Dirichlet
boundary condition at zero are bounded at the infinity for $\lambda \notin
\Xi .$ At the same way we can prove that the linear independent solution $%
v(x,\lambda )$ is bounded at the infinity if $\lambda \notin \Upsilon $ $%
(\dim \Upsilon =0\ ).$ Consequently results obtained by Stolz \cite{St}
guarantee that the support of singular measure has the zero Hausdorff
dimension. Meanwhile as we have shown above spectral measure gives the zero
weight to any set of zero Hausdorff measure. So the singular spectrum is
absent.$\Box$

{\bf Remark 1.} We could have got rid of the condition (\ref{aus}) and prove
the absence of positive eigenvalues by making use of Hardy inequality for
the equation $Q(x)=\int\limits_x^\infty A(s)e^{-i\lambda s}\overline{Q(s)}ds$
which follows from $Q^{^{\prime }}=-Ae^{-i\lambda x}\overline{Q}$ and $%
Q(\infty )=0 $. Really, it would mean the absence of nonzero eigenvalues for
differential system (\ref{s1}). By (\ref{m1}) we would have the absence of
positive eigenvalues for (\ref{e1}).

{\bf Remark 2.} Conditions of Theorem 1 are often fulfilled for potentials
that oscillate at the infinity (see \cite{Behn1}, \cite{Behn2} and
bibliography there ). We used method different from common ones such as
modified Pr\"ufer transform, $I+Q$ asymptotic integration and so on.

{\bf Remark 3.} If we consider potentials which satisfy the estimate $\left|
q_\varepsilon (x)\right| \leq \frac \varepsilon {x+1}$, then the point
spectrum may occur on $[0,\frac{4\varepsilon ^2}{\pi ^2}]$ (see \cite{vNW}, 
\cite{Rem2} ) for any $\varepsilon>0 $. The situation for potentials
considered in Theorem 1 is different. Really, the von Neumann -Wigner
example $q_{vNW}=8\frac{\sin 2x}{x} + O(\frac{1}{x^2})$ shows that the
condition 
\begin{equation}
\left| \int\limits_x^\infty q_\gamma (s)ds\right| \leq \frac \gamma {x+1}
\label{ppp}
\end{equation}
doesn't guarantee the absence of the positive eigenvalues for $\gamma $
large enough. Meanwhile, for small $\gamma \ (0< \gamma <1/4)$ the singular
spectrum disappears on the whole positive half-line. \footnote{%
It's interesing to find out is the constant $1/4$ optimal or not.}

{\bf Remark 4.} From the proof of the Theorem 1 we can infer that the
asymptotics for generalized eigenfunctions may not be true on the set of
zero Hausdorff dimension only. And this statement holds even without the
constraint (\ref{aus}). In \cite{RS} the following simple statement is proved

{\sl If }$q(x)${\sl \ is continuous function which admits the representation 
} $q=\frac{\partial W}{\partial x},\ W\in L^1(1,\infty )${\sl \ then equation%
} $-\varphi ^{^{\prime \prime }}+q\varphi =k^2\varphi ${\sl \ doesn't have
nontrivial square integrable solutions for }$k\neq 0.$ {\sl What is more, if 
}$\left| W(x)\right| \leq C|x|^{-1-\varepsilon }${\sl \ then for any $k\neq
0 $ there exists the solution }$\varphi (x,k)${\sl \ of the same equation
which has the following asymptotic} $\left| \varphi (x,k)-e^{-ixk}\right|
\leq C(k) \left| x\right| ^{-\varepsilon }{\ ,x\geq 1,}${\sl where } $%
|C(k)|<C(k_0)${\sl \ for }$\left| k\right| \geq k_0>0.$

We improved this result in power scale in some extent using method from \cite
{Rem2}.

{\bf Theorem 2. } {\it If }$q(x)${\it \ is real-valued function such that
the improper integral }$W(x)=\int\limits_x^\infty q(s)ds${\it \ exists and
satisfies the condition} $W(x)\in L^p(R^{+})\cap L^2(R^{+}),\ (1<p<2),${\it %
\ then for a.e. positive spectral parameter the generalized eigenfunctions
has the following asymptotic }$u(x,\lambda ,\alpha )=C(\lambda ,\alpha )\sin
(x\lambda +\varphi (\lambda ,\alpha ))+\overline{o}(1), where \ u(0,\lambda
,\alpha)=\cos (\alpha ),\ u^{^{\prime }} (0,\lambda ,\alpha )=\sin (\alpha
). $ {\it \ }

Proof.

Let's consider the system (\ref{s1}) with $a(x)=-W(x).\ $Introducing $%
P(x,\lambda )$ by (\ref{red}) and $Q(x,\lambda )=e^{-i\lambda x}P(x)$ we
have the following equation for $Q(x,\lambda ):$

\begin{equation}
Q^{^{\prime }}=-Ae^{-i\lambda x}\overline{Q},  \label{q1}
\end{equation}

where $Q(0,\lambda )=1,A(x)=\frac 12a(\frac x2).$ So we can see that $%
Q(x,\lambda )=1-\int\limits_0^xA(s)\overline{P(s,\lambda )}ds=1-\overline{%
\int\limits_0^xA(s)P(s,\lambda )ds}.\ $Since $P(x,\lambda )$ play the role
of orthogonal polynomials, we can expect that the analogue of Menchoff
theorem for orthonormal systems \cite{Mench} will guarantee the convergence
of the last integral almost everywhere.

But for our purpose it's more convenient not to derive the generalization of
Menchoff results for $P(x,\lambda )$ system but to give the reference to the
very recent results of M.Christ, A.Kiselev \cite{CK} Really, for the
solution of equation (\ref{q1}) we have the following formal series:

\begin{eqnarray*}
Q(x,\lambda ) &=&Q_\infty (\lambda )(1+\int\limits_x^\infty
A(s_1)e^{-i\lambda s_1}ds_1+\ldots \\
&&+\int\limits_x^\infty A(s_1)e^{-i\lambda s_1}\int\limits_{s_1}^\infty
A(s_2)e^{i\lambda s_2}\ldots \int\limits_{s_{j-1}}^\infty
A(s_j)e^{(-1)^j\lambda s_j}ds_j\ldots ds_1+\ldots ).
\end{eqnarray*}

The convergence of this kind of series for a.e. $\lambda $ w.r.t. Lebesgue
measure was established in \cite{CK} for $A(s)\in L^p(R^{+}),\ 1\leq p<2.$
By the methods of this paper we can show that $Q(x,\lambda )$ satisfies the
equation (\ref{q1}) for a.e. $\lambda .$ Thus we have that a.e. $Q(x,\lambda
)$ is bounded at the infinity. Let's consider now the Sturm-Liouville
operator on the half-line with potential $q^{*}(x)=a^{^{\prime }}+a^2$ with
Dirichlet boundary condition at zero. From (\ref{bis}) and (\ref{red}) we
can infer that the generalized eigenfunctions of this equation which satisfy
the Dirichlet condition at zero are bounded at the infinity for a.e.
positive spectral parameter.

It's obvious that $q^{*}(x)$ can be represented in the following form $%
q^{*}(x)=-T^{^{\prime }}+T^2$, where $T=W$. Repeating the same argument for
Dirac-type system (\ref{s1}) with $a(x)=T(x)$ we see, that generalized
eigenfunctions, which satisfy the conditions $\Phi (0)=1,$ $\Phi ^{^{\prime
}}(0,\lambda )+T(0)\Phi (0,\lambda )=0$ are bounded at the infinity as well.
At the same way we can show that the whole transfer matrix of
Sturm-Liouville equation with potential $q^{*}(x)$ is bounded at the
infinity for a.e. positive spectral parameter. But $q^{*}(x)=W^2-W^{^{\prime
}}=W^2+q,$ so it differs from the initial potential $q$ by the $W^2\in
L^1(R^{+})$ term only. It's easy to show, that if the transfer matrix of
S.L. operator is bounded for some $\lambda ,$ so the transfer matrix of
operator obtained by the $L^1(R^{+})$ perturbation is bounded also.
Consequently the transfer matrix of the initial S.L. operator is bounded for
a.e. positive spectral parameter. $\Box $

{\bf Remark 1}. The fulfillment of the conditions of Theorem 2 doesn't mean
that the singular component of spectrum is absent. Moreover the von Neumann-
Wigner potential illustrates that at least the positive eigenvalue may occur.

Now we will show that conditions of Theorem 2 are satisfied under some
assumption imposed on the $\cos $ -transform of potential. From then on we
suppose that $q(x)$ is such that

(A) its $\cos $ -transform $\widehat{q(\omega )}=\lim_{N\rightarrow \infty
}\int\limits_0^Nq(x)\cos (\omega x)dx\ $exists in $L^{1}_{loc}(R^{+})$ sence.

(B) $\frac 2\pi \lim_{N\rightarrow \infty }\int\limits_0^N\widehat{q(\omega )%
}\cos (\omega x)d\omega =q(x)$ in $L^{1}_{loc}(R^{+}).$

{\bf Theorem 3. }{\it If $q(x)$ is such that (A) and (B) are satisfied and $%
\widehat{q(\omega )}=\widehat{q(0)}+\widehat{\varphi (\omega )}$ where $%
\frac{\widehat{\varphi (\left| \omega \right| )}}\omega \in L^{\varepsilon
,2}(R)$ for some positive $\varepsilon $, then the asymptotics from the
Theorem 2 is true$.$}

Proof. Let's consider even infinitely smooth function $\widehat{\chi (\omega
)}$ such that 
\begin{eqnarray*}
\widehat{\chi (\omega )} =\Bigl\{ 
\begin{array}{cc}
1, & |\omega |\leq 1/2; \\ 
0, & |\omega |\geq 1.\Bigr.
\end{array}
\end{eqnarray*}

Then $\widehat{q(\omega )}=\widehat{q(0)}\widehat{\chi (\omega )}+\widehat{%
\psi (\omega )}$ where $\widehat{\psi (\omega )}=\widehat{q(0)}\widehat{%
[1-\chi (\omega )]}+\widehat{\varphi (\omega ).}$ It's easy to see that $%
\frac{\widehat{\psi (\left| \omega \right| )}}\omega \in L^{\varepsilon
,2}(R).$ And it suffices to prove that S.L. operator with potential $\psi
(x)=\frac 2\pi \int\limits_0^\infty \widehat{\psi (\omega )}\cos (\omega
x)d\omega $ has the transfer matrix bounded at the infinity for a.e.
positive spectral parameter. Really, the initial potential $q(x)=\psi
(x)+\chi (x)\widehat{q(0)}$ where $\chi (x)=\frac 2\pi \int\limits_0^\infty 
\widehat{\chi (\omega )}\cos (\omega x)d\omega =\frac 1\pi
\int\limits_{-\infty }^\infty \widehat{\chi (\omega )}\cos (\omega x)d\omega
\in L^1(R^{+}).$ So the transfer matrix of initial operator would be bounded
at the infinity for a.e. positive spectral parameter.

Consider $\widehat{L(\omega )}=\frac{\widehat{\psi (\left| \omega \right| )}}%
\omega .$ 
\begin{eqnarray*}
W(x) &=&\lim_{N\rightarrow \infty }\int\limits_x^N\psi (s)ds=\frac 2\pi
\lim_{N\rightarrow \infty }\int\limits_x^N\left\{ \lim_{T\rightarrow \infty
}\int\limits_0^T\widehat{\psi (\omega )}\cos (\omega s)d\omega \right\} ds=
\\
&=&\frac 2\pi \lim_{N\rightarrow \infty }\lim_{T\rightarrow \infty
}\int\limits_0^T\widehat{\psi (\omega )}\left\{ \int\limits_x^N\cos (\omega
s)ds\right\} d\omega =\frac 1\pi \lim_{N\rightarrow \infty
}\lim_{T\rightarrow \infty }\int\limits_{-T}^T\widehat{L(\omega )}(\sin
(N\omega )-\sin (x\omega ))d\omega = \\
&=&-\sqrt{\frac 2\pi} \Im \widehat{L(x )},
\end{eqnarray*}

where $L(x)$ is from $L^p(R^{+})$ (for some $p<2$ ) as the Fourier transform
of $L^{\varepsilon ,2}(R)$ function and so the arguments from the Theorem 2
are applicable. $\Box$

We see that roughly speaking these conditions are fulfilled if $\widehat{%
q(\omega )}$ is relatively smooth near the zero and admits some bounds at
the infinity.

{\bf Remark 1}. Local condition in zero of Theorem 3 is satisfied if $%
\widehat{q(\omega )}$ is from $W^{1+\varepsilon ,2}(0,\delta )$ for some
positive $\delta >0.$

{\bf Remark 2. }In paper \cite{Den} it was considered the dependence of
absence of singular component on certain interval on the local smoothness of
Fourier transform.

{\bf Remark 3. }From the corollary and method used in theorem 3 it follows
that

if $q(x)${\it \ is such that (A) and (B) are satisfied, }$\widehat{q(\omega )%
}=\widehat{q(0)}+\widehat{\varphi (\omega )}${\it \ where }$\left| \widehat{%
\varphi (\left| \omega \right| )}\right| \leq C|\omega |^{1/2+\varepsilon }\ 
${\it \ in the vicinity of zero for some positive }$\varepsilon ${\it \ and }%
$\frac{\widehat{\varphi (\left| \omega \right| )}}{|\omega |+1}\in L^2(R),$%
{\it \ then the essential support of spectral measure of operator is }$%
R^{+}. $\vspace{1cm}

{\bf Section C.}

In this section we will discuss the dependence of spectral measure $\sigma
(\lambda )$ on the coefficient $A(x)$ of system (\ref{sys3}).

In fact function $A(x)$ plays the role of sequence $a_n$ for polynomials
orthogonal on the unit circle (see \cite{Ger} ).

We will see that for the Dirac-type systems the situation is not so simple.
The basic reason is the possible oscillation of $A(x)$. The following Lemma
is true

{\bf Lemma. }If measurable bounded function $A(x)$ is such that 
\[
\int\limits_x^\infty e^{-s}A(s)ds=\overline{o}(e^{-x}),\
A(x)e^x\int\limits_x^\infty A(s)e^{-s}ds\in L_1(R^{+}) 
\]
then conditions (1)-(4) from section A are satisfied.

Proof. Consider system (\ref{sys3}) with $\lambda =i$. If $P=e^{-x}Q$ then
we have

\begin{equation}
\begin{array}{ccc}
Q^{^{\prime }} &=&-Ae^xP_{*} \\
P_{*}^{^{\prime }} &=&-Ae^{-x}Q  
\end{array}
\label{ff1}
\end{equation}

Consequently 
\begin{eqnarray*}
P_{*}(x,i)
&=&1-\int\limits_0^xA(s)e^{-s}ds+\int\limits_0^xA(s)e^sP_{*}(s,i)\int%
\limits_s^\infty A(\xi )e^{-\xi }d\xi ds-\int\limits_x^\infty A(\xi )e^{-\xi
}d\xi \int\limits_0^xA(s)e^sP_{*}(s,i)ds \\
&&
\end{eqnarray*}

And now it suffices to use the standard argument. Let $M_n=\max_{x\in
[0,n]}|P_{*}(x,i)|=|P_{*}(x_n,i)|$

So $M_n\leq 1+C+M_n\int\limits_0^\infty |A(s)|e^s\left| \int\limits_s^\infty
A(\xi )e^{-\xi }d\xi \right| ds+\overline{o}(1)M_n$

If the whole integral in the last formula is less then $1$ then $M_n$ is
bounded. Otherwise we should start to solve the equations (\ref{ff1}) not
from zero but from some other point $x_0$ for which this condition is
satisfied.

{\bf Example. }$A(x)=(x^2+1)^{-\alpha }\sin (x^\beta )$ where $\alpha ,\beta
>0.$

One can easily verify that conditions of Lemma are satisfied if $2\alpha
+\beta /2>1$. Meanwhile $A(x)\in L_2(R^{+})$ if and only if $\alpha >1/4$.
Nevertheless for nonpositive $A(x)$ with bounded derivative the condition $%
A(x)\in L^2(R^{+})$ is necessary for (1)-(4) from Section A to be true.

{\bf Proposition. }If one of the conditions (1)-(4) is true, $A(x)\leq 0$
and $A^{^{\prime }}(x)$ is bounded then $A(x)$ is from $L^2(R^{+}).$

Proof.

Really, since $A(x)\leq 0$ both $P$ and $Q$ are not less then 1.
Consequently if one of (1)-(4) holds then $P_{*}(x,i)$ is bounded and as it
follows from (\ref{ff1}) $\int\limits_0^x|A(s)|e^s\int\limits_s^x\left|
A(\xi )\right| e^{-\xi }d\xi ds$ is bounded as well. But we have the
inequality

\[
\int\limits_0^x|A(s)|e^s\int\limits_s^x\left| A(\xi )\right| e^{-\xi }d\xi
ds\geq e^{-1}\int\limits_0^{x-1}|A(s)|\int\limits_s^{s+1}\left| A(\xi
)\right| d\xi ds\geq C\int\limits_0^{x-1}\left| A(s)\right| ^2ds. 
\]

which concludes the proof of proposition. The latter inequality follows from
the boundedness of $A^{^{\prime }}(x).$ $\Box$

Function $A(x)$ from the example above with $\alpha =1/4,\ 1<\beta \leq 3/2\ 
$ satisfies the conditions of Lemma (consequently (1)-(4) holds ), has
bounded derivative but is not from $L^2(R^{+}).$ The explanation is that
this function is not nonpositive.

We would like to conclude the paper with two open problems the first of
which is much more difficult then the second one.

{\bf Open problems}.

1. Prove that Theorem 2 holds for $W\in L^2(R^{+})$.

2. Prove that the presence of a.c. component on the half-line pertains to
those potentials which Fourier transform is from $L^{2}$ near the zero.
Specifically, if $q$ admits the Fourier transform $\widehat{q}$ such that $%
\widehat{q}\in L^{2}_{loc}(R)$ and $\frac{\widehat{q}}{|\omega |+1}$$\in
L^2(R), $ then the a.c. part of the spectrum fills the whole positive
half-line. This conjecture seems reasonable at least with some additional
constraints since we can represent $\widehat{q}=\widehat{q_1}+\widehat{q_{2}}$. Where the $\widehat{q_1}$
is localized near the zero and is from $L^2$ so the methods of paper \cite
{Den} works. The other function $\widehat{q_{2}}$ is such that Theorem 3 can be
applied.

Acknowledgment. Author is grateful to C.Remling for attention to this work.

e-mail address: saden@cs.msu.su

\end{document}